\documentclass[twocolumn,preprintnumbers,amsmath,amssymb,aps,pre,reprint,longbibliography]{revtex4-2}
\usepackage{graphicx}
\usepackage{xcolor}
\begin{document}

\title{Reversible to Irreversible Transitions in Pattern-Forming Systems with Cyclic Interactions}
\author{
C. Reichhardt and C. J. O. Reichhardt 
} 
\affiliation{
Theoretical Division and Center for Nonlinear Studies,
Los Alamos National Laboratory, Los Alamos, New Mexico 87545, USA
}

\date{\today}

\begin{abstract}
Transitions from reversible to irreversible or fluctuating states above a critical density and shear amplitude have been extensively studied in non-thermal cyclically sheared suspensions and amorphous solids.
Here, we propose that the same type of reversible to irreversible transition occurs for a system of particles with competing short-range attraction and long-range repulsion, which can form crystals, stripes, and bubbles as the ratio of attraction to repulsion varies.
By oscillating the strength of the attractive part of the potential, we find that the system can organize into either time-periodic states consisting of nondiffusive complex closed orbits, or into a diffusive fluctuating state.
A critical point separates these states as a function of the maximum strength of the attraction, oscillation frequency, and particle density. We also find a re-entrant behavior of the reversible state as a function of the strength of the attraction and the oscillation frequency.
\end{abstract}

\maketitle

{\it Introduction---}
A non-thermal suspension of colloidal particles under cyclic shear
organizes
after a number of oscillations into a steady state that can be either
reversible, where the particles return to the same positions after
each cycle, or irreversible or fluctuating,
where the particles undergo diffusion over
repeated cycles \cite{Pine05,Corte08,Reichhardt23}.
Above a critical shear amplitude at fixed density or a critical density
at fixed shear amplitude,
the system remains in an irreversible state, and this
critical point is accompanied by
a divergence in the number of cycles required for the establishment of
a steady state
\cite{Corte08,Menon09,Weijs15,Tjhung15,Lei19a,Wilken20,Wang22,Reichhardt23}.
The reversible-irreversible transition is an example of a nonequilibrium
absorbing phase transition,
and when particle collisions are absent in the reversible state,
the transition is referred to as 
random organization \cite{Corte08,Hexner17,Reichhardt23}. 
Similar reversible (R) to irreversible (IR) transitions have also been
observed for cyclic shearing of strongly interacting systems,
such as amorphous solids,
where in the reversible state, the particles return to their
original locations after executing
complex two-dimensional orbits that may span one or multiple shear cycles
\cite{Regev13,Keim14,Regev15,Priezjev16,Jana17,Lavrentovich17,Lindeman21,Keim21,Khirallah21}. 
Periodically sheared systems that exhibit R-IR transitions 
additionally show a variety of memory effects
\cite{Paulsen14,Fiocco14,Keim19,Mungan19,Paulsen25}.
R-IR transitions have also
been studied for periodically driven systems moving
through quenched disorder, such as superconducting vortices
with random pinning \cite{Mangan08,Maegochi19} 
and colloidal particles in obstacle arrays \cite{Reichhardt22,Minogue24}. 

An open question is whether similar R-IR
transitions can occur for other types
of nonequilibrium systems in the absence of
periodic shear but the presence of some other type of cyclic driving,
such as a periodic oscillation of the pairwise interaction potential
between the particles.
There are a variety of systems that can be effectively
modeled with competing short-range attraction and long-range repulsion (SALR),
which form crystal, stripe, labyrinth, void lattice, and bubble
states as a function of the ratio of attraction to repulsion or
the particle density
\cite{Reichhardt04,Nelissen05,Liu08,Reichhardt10,Xu11,CostaCampos13,Liu19,Wang21,Xu21,AlHarraq22,Hooshanginejad24}.
In a recent experimental realization of a SALR system
\cite{Hooshanginejad24},
Hooshanginejad {\it et al.} could continuously vary
the ratio of competing magnetic and capillary
interactions between particles by applying a magnetic field. 
In SALR systems of this type, it would be possible to cycle
the ratio of the attraction to repulsion
as a function of time, and see whether 
the particles settle into a steady state
where they return to the same positions after each cycle or
remain in a fluctuating or liquid state.

Here, we consider a two-dimensional (2D) system of
interacting particles with
competing short-range attraction and long-range repulsion,
where we cycle the amplitude of the attractive interaction.
For a fixed oscillation frequency and
particle density,
the system always forms a reversible state when the maximum attraction
amplitude is weak.
We find that
there is a critical attraction amplitude
above which the system becomes irreversible and displays
diffusive behavior similar to that
found in cyclically sheared systems.
We also find critical oscillation frequencies and
particle densities where the
system organizes into a reversible state, and near this
R-IR transition, the time scale for reaching a steady state diverges as
a power law.
In the reversible state there can be
complex orbits where groups of particles traverse closed loops around
patches of particles that remain immobile.
At higher densities,
dynamical lace-like patterns of motion can appear.
The time-periodic reversible states can be
regarded as examples of a classical 
discrete time crystal  \cite{Libal20,Yao20,Ernst22,Zhao25}.
Our results show that R-IR transitions can be realized by
oscillating the interaction potential,
rather than by applying a cyclic shear,
which could open new ways to study memory and dynamic pattern formation.

\begin{figure}
  \includegraphics[width=0.98\columnwidth]{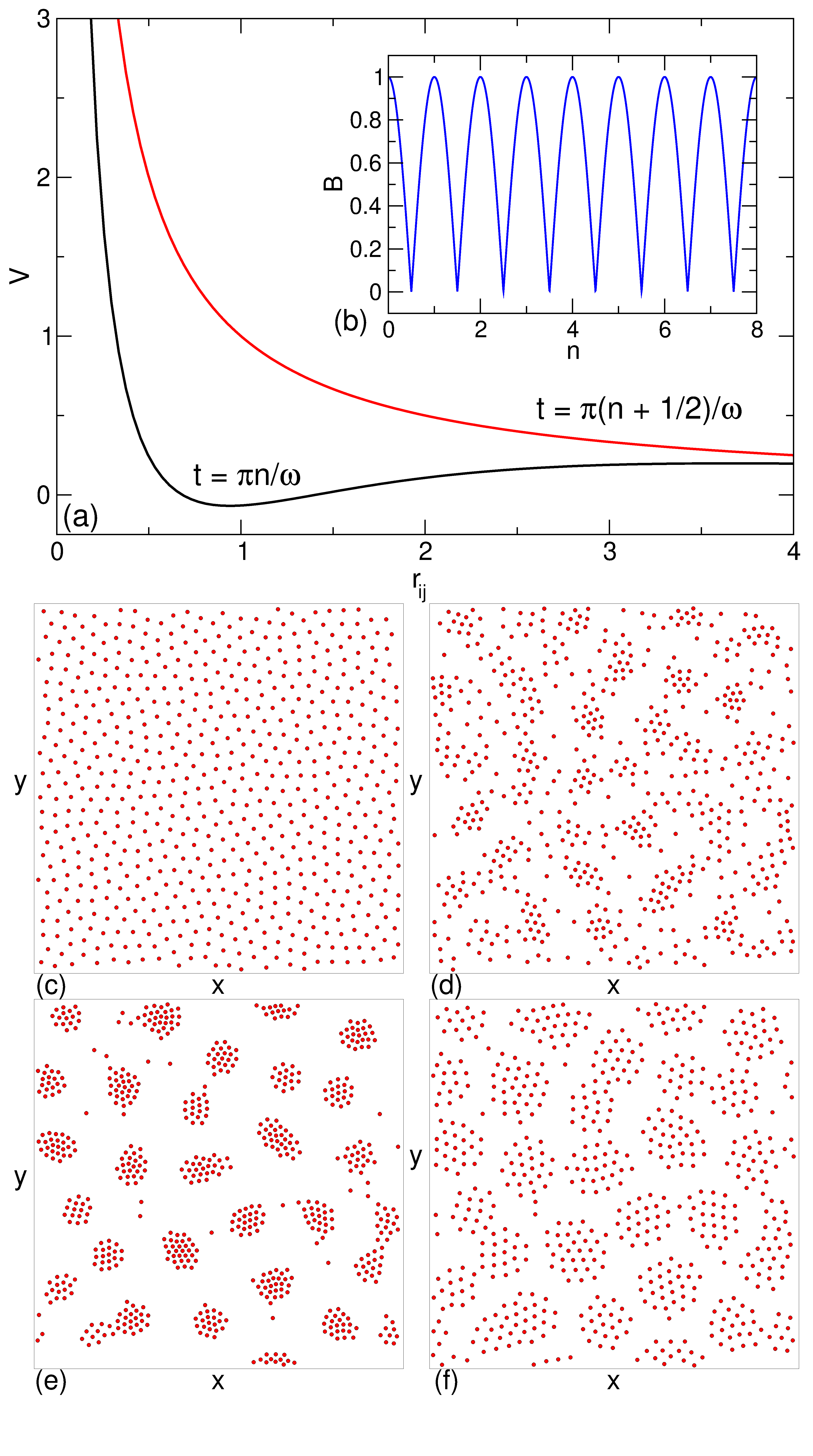}
\caption{(a) The SALR interaction potential
$V(r_{ij},t)=1/r_{ij} - B(t)\exp(-r_{ij})$
for $B_{\rm max}=2.9$ and $\omega=1.25\times 10^{-4}$
at two different times,
$t=\pi n/\omega$ (black) where $B(t)=B_{\rm max}$ and
$t=\pi(n+1/2)/\omega$ (red) where $B(t)=0$ and the interaction is
purely repulsive.
(b) The corresponding short range interaction strength
$B(t) = B_{\max}|\cos(\omega t)|$ plotted against cycle number $n$.
(c-f) Particle configurations for a system with $B_{\rm max}=2.9$, $\rho=0.5$
and $\omega=1.25\times 10^{-4}$ during
cycle $n=20$.
(c) Uniform state at $B/B_{\rm max} = 0.0$.
(d) Beginning of bubble formation at $B/B_{\rm max} = 0.75$ for increasing $B$.
(e) Fully formed bubbles at $B/B_{\rm max} = 1.0$.
(f) The bubble structure expands back toward a uniform state
at $B/B_{\rm max} = 0.5$ for decreasing $B$.
}
\label{fig:1}
\end{figure}

{\it Simulation---}
We model a 2D system of size $L \times L$
with periodic boundary conditions in the
$x$ and $y$ directions.
The system contains $N$ particles with a density $\rho=N/L^2$, and
we take $L=36$.
The particles have a time-dependent SALR interaction potential of the form
\begin{equation} V(r_{ij},t) = \frac{1}{r_{ij}} - B(t)\exp(-r_{ij}) \ . \end{equation}
The distance between particles $i$ and $j$ is $r_{ij}=|{\bf r}_i-{\bf r}_j|$,
and $B$ is the strength of the attraction.
The long-range repulsion favors the formation of a uniform triangular lattice, while the attractive term favors bubble formation.
For increasing $B$ at a fixed density, crystal, stripe, and bubble phases
appear, while for increasing $\rho$ at fixed $B$,
crystal, bubble, stripe, and void lattice phases emerge
\cite{Reichhardt10,Reichhardt24a}. 
The particles are initially placed in random, non-overlapping positions,
and their dynamics evolve according to
the overdamped equation
\begin{equation} \eta \frac{d {\bf r}_{i}}{dt} = -\sum^{N}_{j \neq i} \nabla V(r_{ij}) \end{equation}
The damping term $\eta=1.0$, and we
employ a Lekner summation method for
computational efficiency in calculating the long-range
repulsive term \cite{Lekner91,GronbechJensen97a}.
The strength of the attractive term
oscillates with time between $B=0.0$ and $B=B_{\rm max}$,
$B(t) = B_{\max}|\cos(\omega t)|$.
For convenience, we designate one cycle as
the progression of $B$ from $B=B_{\max}$ to $B=0.0$ and back,
so that there are two cycles per
period $\tau=2\pi/\omega$.
In Fig.~\ref{fig:1}(a), we plot $V(r_{ij},t)$ for a system with
$B_{\max} = 2.9$ and $\omega=1.25\times 10^{-4}$ at $t=\pi n/\omega$, where
$B = 0.0$ and the interaction is purely repulsive, and
at $t=\pi(n+1/2)/\omega$, where $B=B_{\max}$ and the attractive and repulsive
interaction terms are competing.
Figure~\ref{fig:1}(b) shows $B(t)$ as a function of cycle number $n$.

To quantify the behavior, we measure
the cumulative displacements of the particles over time, 
$d(n) = \sum^N_{i}|{\bf r}_i(t=n) - {\bf r}_{i}(t=0)|$, and
the net displacement of the particles after a single cycle,
$R(n) = \sum^{N}_{i}[{\bf r}_{i}(t=n) - {\bf r}_{i}(t=n-1)]$. 
If the system is in a fluctuating state,
$d(n)$ increases monotonically with time
and $R(n)$ is finite,
while in reversible states,
$d(n)$ saturates and $R(n)\approx 0$.
These are  the same measures used in 
periodically sheared systems to
detect R-IR transitions \cite{Pine05,Corte08,Reichhardt23}.

{\it Results---}
In Fig.~\ref{fig:1}(c) we plot
the particle configurations during cycle $n=20$
for a system with $\rho = 0.5$, $B_{\rm max}=2.9$, and
$\omega = 1.25 \times 10^{-4}$
in the $B/B_{\rm max} = 0$ portion
of the cycle, where a uniform and partially ordered structure appears.
During the increasing $B$ portion of the cycle
at $B/B_{\rm max} = 0.75$ in Fig.~\ref{fig:1}(d), the particles are beginning to
aggregate into bubbles.
Figure~\ref{fig:1}(e) shows that well defined bubbles are present
when $B/B_{\rm max} = 1.0$.
As $B$ decreases again,
Fig.~\ref{fig:1}(f) indicates that
at $B/B_\text{max} = 0.5$
the clumps are expanding back toward a uniformly dense state.
We note that for $\rho = 0.5$ at constant $B$,
the system forms a crystal for $B \leq 2.0$,
stripes for $2.0 < B < 2.3$, and clumps for $B \geq 2.3$.

\begin{figure}
  \includegraphics[width=\columnwidth]{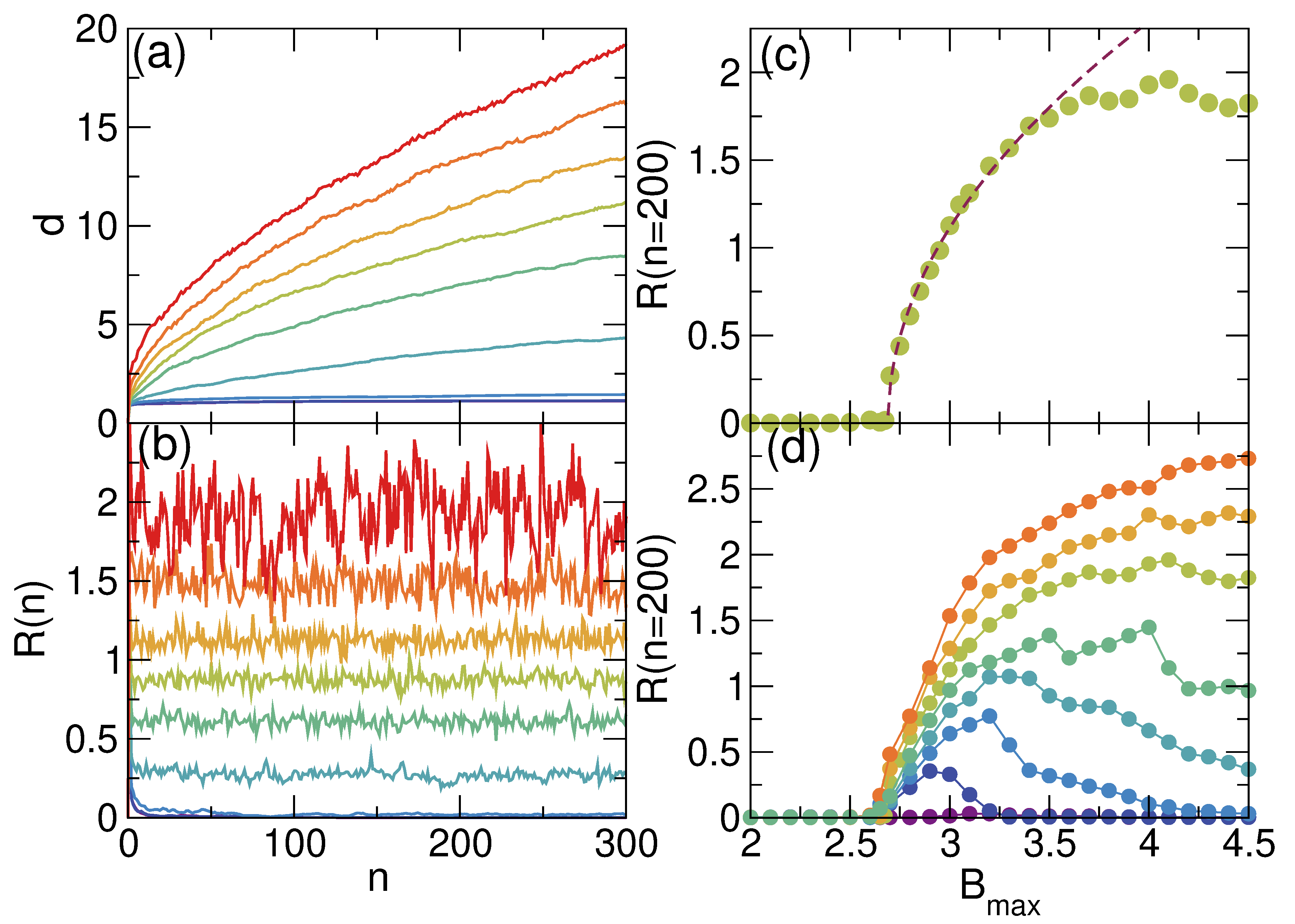}
\caption{(a) Cumulative displacements $d(n)$ vs $n$ for a system with
$\rho=0.5$ and $\omega=1.25\times 10^{-4}$ at  
$B_{\rm max} = 2.0$, 2.4, 2.6, 2.7, 2.8, 2.9, 3.0, 3.2, and $4.0$,
from bottom to top.
(b) The corresponding $R(n)$ vs $n$.
The behavior is reversible for $B_{\rm max} < 2.7$
and irreversible for $B_{\rm max} \geq 2.7$.
(c) $R$ after $n=200$ cycles vs
$B_{\rm max}$ for the same system,
showing a critical transition near
$B_{\rm max} = 2.65$.
The dashed line is a fit to
$R \propto (B_{\rm max} - B_{\rm max}^c)^\beta$ with
$B_{\rm max}^c = 2.69$ and $\beta = 1/2$.
(d) $R$ after $n=200$ cycles vs $B_\text{max}$ for
$\rho = 0.2$, 0.225, 0.25, 0.275, 0.3, 0.35, 0.4, 0.5, 0.6, and $0.8$,
from bottom to top. The lowest three values of $\rho$ give $R=0$
for all $B_{\rm max}$.
}
\label{fig:2}
\end{figure}

In Fig.~\ref{fig:2}(a,b),
we plot the cumulative displacement $d(n)$
and displacement per cycle $R(n)$, respectively,
versus $n$ for a system with $\rho = 0.5$
and $\omega=1.25\times 10^{-4}$
at $B_{\rm max} = 2.0$, 2.4, 2.6, 2.7, 2.8, 2.9, 3.0, 3.2, and $4.0$.
For $B_{\rm max} < 2.7$, $d(n)$ saturates to a constant value
and $R(n)$ is close to zero since
the system organizes into a reversible state.
In contrast, for $B_{\rm max} \geq 2.7$,
$d(n)$ increases monotonically with $n$ as $n^{1/2}$, the
expected behavior for Brownian motion,
while $R(n)$ has a finite value,
indicating that the particles do not return to their
original positions after each cycle.
This behavior of $d(n)$ and $R(n)$ is the same as what is found at the
R-IR transition
in periodically sheared colloidal suspensions \cite{Pine05,Corte08}.
In Fig.~\ref{fig:2}(c), we show the value of $R$ after
$n=200$ cycles versus
$B_{\rm max}$ for the same system,
where we observe a critical
amplitude $B_{\rm max}^c \approx 2.65$
for the transition to an irreversible state.
The dashed line is a fit to
$R \propto (B_{\rm max} - B_{\rm max}^c)^\beta$
with $B_{\rm max}^c = 2.69$ and $\beta = 1/2$.
This is the same exponent found 
by Corte {\it et al.} \cite{Corte08} for the R-IR transition,
and is consistent with the
directed percolation universality class \cite{Hinrichsen00,Reichhardt23}.
Figure~\ref{fig:2}(d) shows $R$ after $n=200$ cycles versus $B_{\rm max}$ for 
$\rho = 0.2$, 0.225, 0.25, 0.275, 0.3, 0.35, 0.4, 0.5, 0.6, and $0.8$.
For $\rho \leq 0.2$ and $B_{\rm max} < 2.6$, the behavior is reversible,
indicating that there is both a critical amplitude
$B_{\rm max}^c$ and
a critical density $\rho_c$ for
observing the R-IR transition, similar to what is found
in other systems \cite{Reichhardt23}.
For $0.225 \leq \rho \leq 0.35$, we find a reentrant R-IR transition in
which
$R$ is zero for
$B_{\rm max} < 2.6$, increases to a nonzero value as $B_{\rm max}$ increases,
but then drops back to a lower value or to zero at
higher $B_{\rm max}$.
For densities within this range,
R-IR and IR-R transitions occur at two different critical amplitudes
$B_{\rm max}^{c_1, c_2}$.

\begin{figure}
  \includegraphics[width=\columnwidth]{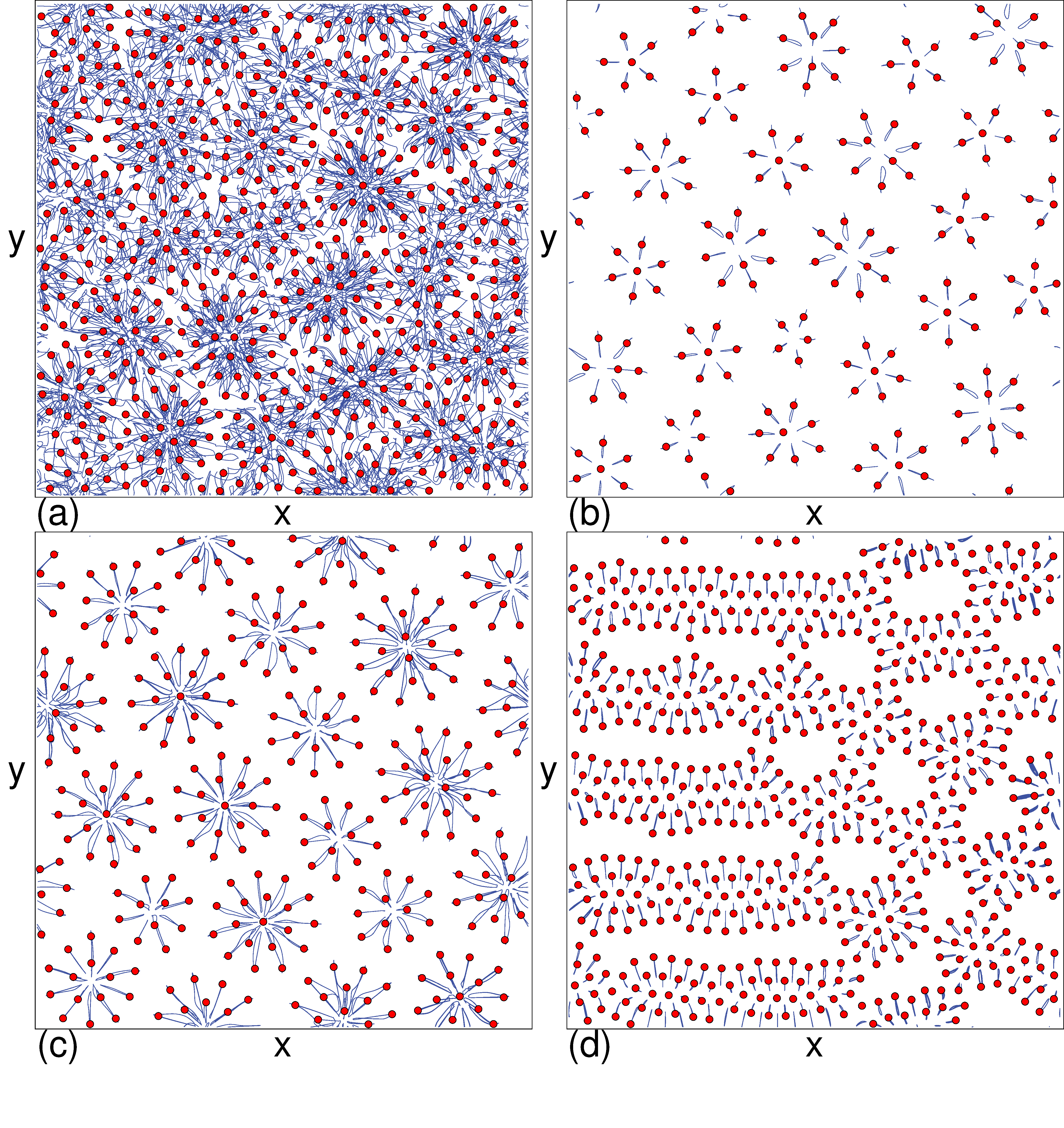}
\caption{Particle configurations (red) and trajectories (blue)
during $n=5$ cycles in the steady state.
(a) Disordered flow in the irreversible state at
$\rho = 0.5$, $B_{\rm max} = 2.9$,
and $\omega=1.25\times 10^{-4}$.
(b) Reversible state at
$\rho=0.15$, $B_{\rm max}=2.9$,
and $\omega=1.25\times 10^{-4}$.
(c) Reentrant reversible state at
$\rho = 0.25$, $B_{\rm max}=4.0$,
and $\omega=1.25\times 10^{-4}$.
(d) A stripe-like reversible state at
$\rho = 0.5$ and $B_{\rm max} = 3.2$
for a higher frequency of $\omega = 8.0\times 10^{-4}$.
}
\label{fig:3}
\end{figure}

In Fig.~\ref{fig:3}(a) we illustrate
the particle locations and trajectories during $n=5$ cycles
in the steady state for a system with
$B_{\rm max} = 2.9$, $\rho = 0.5$, and $\omega=1.25\times 10^{-4}$.
The motion is disordered and the behavior is irreversible.
At $\rho=0.15$ in the same system, shown in Fig.~\ref{fig:3}(b),
we find a reversible state where long-time diffusion is absent
and the trajectories are more ordered.
In general, particles follow the exact same path during each cycle
in the reversible regime,
but we find some cases
where individual particles return to their original locations
after multiple cycles
or even undergo an exchange with particles in the same cluster;
however, there is no long-time diffusion.
The reversible state in Fig.~\ref{fig:3}(b)
can be viewed as an example of a dissipative time crystal,
since the overall structure repeats as a function of time \cite{Libal20,Yao20}.
In Fig.~\ref{fig:3}(c), we show the
$B_{\rm max}=4.0$ reentrant reversible state
that appears for large $B_{\rm max}$ in the
$\rho = 0.25$ system from Fig.~\ref{fig:2}(d).
At this density and oscillation frequency, the repulsion dominates for
$B_{\rm max} < 2.6$ and the system remains in
a uniform distorted crystal state.
When $B_{\rm max}$ or $\rho$ increase
enough for the system to reach a clump state during the
$B/B_{\rm max} \approx 1$ portion of the cycle,
an irreversible state can emerge in which
the system cycles between uniform and clump
states. The irreversibility arises if the clumps that form are composed
of a different set of particles on each cycle
due to mixing that occurs as the system passes in and out
of the uniform state, meaning that some particles effectively jump from
one clump to another between cycles.
At higher $B_{\rm max}$, the clump structure 
becomes so compact that particles inside the clumps
do not have time to fully expand into the uniform crystal
state during the decreasing $B$ portion of the cycle,
so each particle becomes confined permanently to a particular clump,
as shown for the reentrant reversible state in Fig.~\ref{fig:3}(c).
The R-IR transition 
is also a function of the oscillation frequency $\omega$,
and as the frequency varies, a quasi-pattern can emerge in
the reversible state,
such as the reversible stripe state shown in
Fig.~\ref{fig:3}(d) at $B_{\rm max} = 3.2$ and $\omega = 8.0 \times 10^{-4}$.
If we fix $B=3.2$ and do not oscillate the interaction potential,
the system forms a compact bubble state.
This indicates that introduction of dynamical oscillation can generate
dynamic patterns that would be unstable in a static system.

\begin{figure}
  \includegraphics[width=\columnwidth, trim = 20 20 20 20, clip]{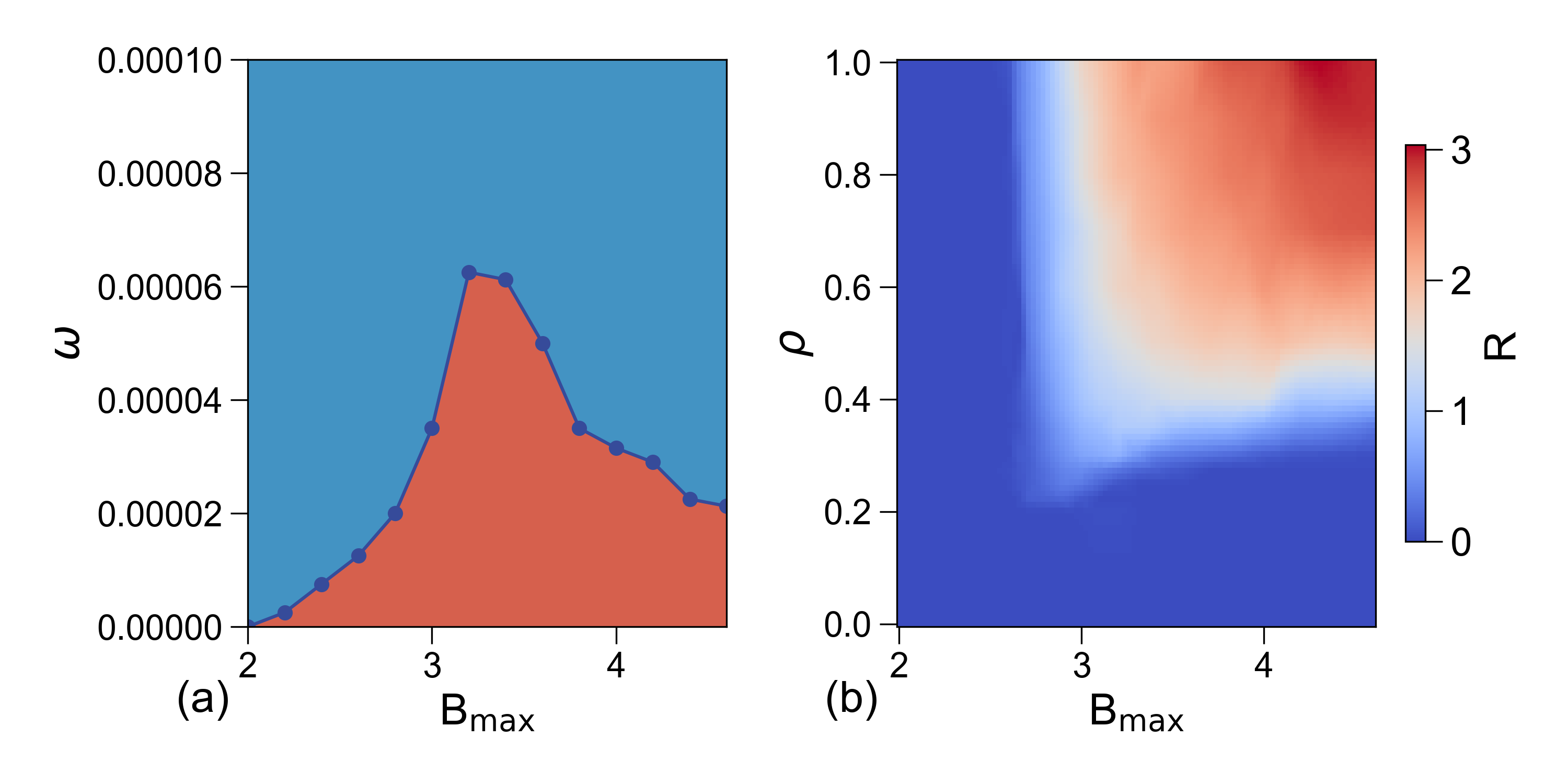}
\caption{(a) Boundary between reversible (blue) and irreversible (red) regimes
as a function of $\omega$ vs $B_{\rm max}$ for $\rho=0.5$.  
When $B_{\rm max} < 2.0$ and/or at high $\omega$,
the system is in a reversible uniform state.
As a function of $B_{\rm max}$ there is a reentrant transition to the
reversible state, and the irreversible state reaches its greatest extent
near $B_{\rm max} = 3.2$.
(b) A heat map of $R$ after $n=200$ cycles
as a function of $\rho$ vs $B_{\rm max}$
in a system with $\omega=1.25\times 10^{-4}$ showing reversible (blue) and
irreversible (red) regimes.
}
\label{fig:4}
\end{figure}

In Fig.~\ref{fig:4}(a) we plot the R-IR boundary as a function of
$\omega$ versus $B_{\rm max}$ at $\rho=0.5$.
There is a reversible state whenever
$\omega$ is large enough that the particles are able to travel a significant
distance during an oscillation cycle.
For $B_{\rm max} < 2.0$, where the attraction is weak, the reversible
configuration is a uniform lattice, and no irreversible behavior occurs
even for arbitrarily low $\omega$.
For $B_{\rm max} = 2.6$,
irreversible behavior appears when
$\omega < 1.25 \times 10^{-4}$, in agreement with the results
in Fig.~\ref{fig:2}(d).
As $B_{\rm max}$ increases, the reversible transition shifts to larger
values of $\omega$
and reaches a maximum in $\omega$ at $B_{\rm max} = 3.2$,
where at high frequencies, the system forms a reversible stripe state
of the type shown in Fig.~\ref{fig:3}(d).
For $B_{\rm max} > 3.2$,
the reversible transition drops to lower
values of $\omega$ since,
as the particles become more strongly compressed
by the attractive portion of the interaction potential,
they require more time to expand into
a uniform state where irreversible mixing can occur.
The re-entrant behavior of the reversible state as a function
of $\omega$ is also consistent
with the re-entrant behavior observed for
varied $B_{\rm max}$ at different $\rho$, as shown in Fig.~\ref{fig:2}(d). 

Figure~\ref{fig:4}(b) shows a heat map of $R$ after $n=200$ cycles
as a function of $\rho$ versus $B_{\rm max}$
in a system with $\omega=1.25\times 10^{-4}$.
At low $\rho$ and low $B_{\rm max}$, the system is in a reversible state.
We observe reentrant reversible behavior at higher $B_{\rm max}$ when $\rho$
is not too large.
It is possible that additional phases are present
within the irreversible regime,
such as a state in which
particles within a cluster exchange irreversibly
but do not leave the cluster,
so that there is no long-time diffusion.
In general, irreversible behavior arises
whenever the combination of $B_{\rm max}$,
$\omega$,
and $\rho$ is such that the
system can pass through one of
the equilibrium phase boundaries,
such as bubble to stripe or bubble to uniform,
slowly enough that the particles have time to mix.

{\it Summary---} We have proposed a
new type of oscillatory interaction driven non-thermal system
that exhibits transitions from reversible states,
where there is no long-term diffusion,
to irreversible or fluctuating states with finite diffusion.
Previous work on reversible-to-irreversible transitions
focused on suspensions of particles or amorphous solids
subjected to oscillatory shear.
Here we consider
a pattern-forming system with competing short-range attraction and
long-range repulsion,
which exhibits uniform crystal, stripe, and bubble
states as the ratio of attraction to repulsion is varied.
When we oscillate
the attractive term, reversible or irreversible behavior appears
depending on the amplitude and frequency of the oscillation and the
particle density.
In the reversible states, it is possible for the
particles to undergo complex orbits.
The transition to irreversible motion
occurs when the system has enough time to relax from the bubble or stripe
state into a uniform crystal state during the oscillation period.
We show that the reversible motion state can be reentrant
as a function of oscillation frequency or maximum attraction amplitude.
The reversible states can
take the form of dynamic patterns that are not stable in a
non-driven system, 
and these states can be
viewed as examples of discrete time crystals.
Our results should be general to a broader class of
pattern-forming systems with
effective competing interactions that can be oscillated as a function of time.

\begin{acknowledgements}
We gratefully acknowledge the support of the U.S. Department of
Energy through the LANL/LDRD program for this work.
This work was supported by the US Department of Energy through
the Los Alamos National Laboratory.  Los Alamos National Laboratory is
operated by Triad National Security, LLC, for the National Nuclear Security
Administration of the U. S. Department of Energy (Contract No. 892333218NCA000001).
\end{acknowledgements}

\bibliography{mybib}

\end{document}